# Towards an AI Buddy for every University Student? Exploring Students' Experiences, Attitudes and Motivations towards AI and AI-based Study Companions

Judit Martínez Moreno[1,2]*, Markus Christen[3], Abraham Bernstein[3,4]

[1] Institute of Education, University of Zurich, Freiestrasse 36, CH-8032 Zurich

[2] Centre for Education and Digital Transformation, Zurich University of Teacher Education, Lagerstrasse 2, 8090 Zurich

[3] UZH Digital Society Initiative, University of Zurich, Rämistrasse 69, CH-8001 Zürich

[4] Department of Informatics, University of Zurich, Binzmühlestrasse 14, CH-8050 Zürich

**Abstract**

Despite the widespread integration of generative artificial intelligence (GenAI) tools in higher education, there is limited empirical insight into students' experiences, competences, and readiness to adopt personalized AI companions. To address this gap, this study investigates three key questions: (RQ1) What are students' prior experiences with AI tools, their perceived digital and AI-related competences, and their interest in emerging technologies?; (RQ2) How do students perceive a hypothetical "AI Buddy" (a digital companion designed to support students throughout their academic journey) including adoption, benefits, and concerns?; (RQ3) How does students' willingness to adopt an AI Buddy relate to motivations for engaging in traditional academic activities? Based on a survey of 926 students at a Swiss university, students revealed widespread prior use of AI, primarily for text-based and productivity tasks, with moderate self-assessed digital competence. Students expressed strong enthusiasm for adopting an AI Buddy, valuing its potential for time efficiency, personalized academic support, and study organization, but expressed significant concerns about data privacy and over-reliance. A weak negative correlation emerged between AI Buddy adoption willingness and motivations for attending lectures or using library resources, while social and collaborative motivations remained unaffected. These findings suggest that AI Buddies may partially replace information-seeking behaviours but preserve the social fabric of university life. This study provides practical recommendations including the need for robust privacy protections and critical engagement strategies to ensure AI Buddies enhance, rather than undermine, the academic and communal value of higher education.

---

* Corresponding author. E-mail address: martinez.moreno.judit@gmail.com
Co-author E-mail addresses: christen@ifi.uzh.ch / bernstein@ifi.uzh.ch

**1. Introduction**

The integration of artificial intelligence (AI) into higher education has undergone a significant transformation since its first mentions in the mid-20th century. Early efforts, such as the development of intelligent tutoring systems (ITS) in the 1970s, aimed to replicate human tutoring by providing individualized instruction tailored to each student's pace and understanding (Hofmeister and Ferrara, 1984; Watters, 2023). These systems were developed with rule-based algorithms and were limited by the computational capacity of the time. However, advances in machine learning, natural language processing (NLP), and data analytics have since expanded the role of AI, enabling the creation of sophisticated tools such as personalized learning platforms (Luckin et al., 2016). Research shows that these modern applications of AI can significantly improve learning outcomes by providing real-time feedback, identifying knowledge gaps, and personalizing educational content to meet the diverse needs of students (Zawacki-Richter et al., 2019). Such possibilities open the door to an AI companion or "AI Buddy" for each student, understood as a personalized assistant that accompanies learners across courses and semesters, potentially orchestrating everything from reference management to fostering social connections.

However, despite these promises, successful implementation depends on students' motivational readiness. Adoption research consistently shows that prior experience, perceived competence, and attitudes shape technology adoption (Sergeeva et al., 2025; Wang & Li, 2024). However, currently there is missing evidence on how these factors influence the adoption of an AI companion, as well as potential ripple effects on the engagement in traditional academic activities. To explore these aspects, the current study investigates the following research questions:

1. What are students' prior experiences with AI tools, their perceived digital and AI-related competences, and their interest in emerging technologies?

2. How do students perceive a hypothetical "AI Buddy" (a digital companion designed to support students throughout their academic journey) including adoption, benefits, and concerns?

3. How does students' willingness to adopt an AI Buddy relate to motivations for engaging in traditional academic activities?

The answers to these questions will provide an empirical basis for universities that wish to use AI assistants without compromising the collaborative and intellectual value of face-to-face teaching.

## 2. State of the Art

### 2.1. AI Integration and Implications for Higher Education

Artificial intelligence (AI) has rapidly advanced from specialized applications to widely used systems that can assist in teaching, learning, and administrative procedures in higher education. According to recent studies, generative AI, such as large language models (LLMs) like ChatGPT, has sparked debates about how it can revolutionize conventional teaching methods while offering opportunities for efficiency and personalization (Jensen et al., 2025; McGrath et al., 2025). Some universities in the U.S., such as Columbia, NYU, and MIT, are actively preparing students through AI-focused curricula, highlighting a growing interest in emerging technologies but also addressing disparities in access and readiness (TOI Education, 2025).

However, the rise of AI tools has brought renewed attention to the purpose and value of the physical university environment. As higher education institutions increasingly rely on digital and hybrid modalities, scholars have questioned whether the growing reliance on intelligent systems risks eroding the communal, dialogic, and serendipitous aspects of university life (Biesta, 2009; Ng et al., 2021). While virtual tools can replicate certain cognitive functions, they cannot easily substitute the embodied and social dynamics of in-person seminars, peer interaction, academic community-building, or even personal relationships. In this context, the integration of persistent AI companions invites reflection on how the presence of highly capable AI support tools influences students' motivation to engage with the university as a physical space.

Some of the benefits of using AI tools include possibilities for personalized learning, brainstorming support, and enhanced engagement, with students viewing AI companions as supplements that free cognitive resources for deeper academic pursuits (Chan & Hu, 2023; Rahman et al., 2025; Shahzad et al., 2025). Also, it has been seen that AI chatbots can facilitate companionship, mitigating loneliness and depression through socioemotional interactions (Lai et al., 2025). However, there are some concerns associated to the use of AI, including ethical issues like privacy, bias, and academic dishonesty, accuracy and over-reliance risks, and potential job displacement (Baek et al., 2024; Chan & Tsi, 2024; Yusuf et al., 2024).

Within this dynamic field, the concept of an AI companion or "AI Buddy" represents a pioneering shift. Unlike conventional AI tools, which focus exclusively on academic tasks such as solving math problems or grading essays, an AI Buddy is designed as a multifaceted virtual companion that supports students in a comprehensive manner, accompanying university students throughout their studies. Leveraging advanced natural language processing and predictive models as well as background data about the coursework and the student, an AI Buddy can engage in natural conversations, offer personalized academic advice (e.g., study schedules or resource recommendations), and address socio-emotional needs, such as stress management or motivation improvement (Williamson et al., 2023). For example, these tools can act as a personalized tutor that helps students in their learning processes, by adapting learning tasks to their level of competence or by giving real-time detailed feedback (Chiu et al., 2023), suggesting exercises during exam periods, or connecting students with peer study groups with common interests. This broader scope distinguishes the AI Buddy from traditional tools and positions it as a game-changer in how students experience university life.

However, while some studies have evaluated the impact of specific AI applications, such as chatbots for language learning (Gupta and Chen, 2022) or virtual tutors for STEM subjects (Ng et al., 2021), few have explored the implications students' attitudes about a general AI Buddy and their impact in the university ecosystem.

### 2.2. Students' AI Adoption and Motivational Dynamics in Higher Education

The use of AI has become very common among university students, with recent studies indicating that almost all students have already used AI for education purposes (Digital Education Council Global AI Student Survey, 2024; Chan & Tsi, 2024). Some studies show that students' willingness to adopt AI-based study companions, such as conversational agents or chatbots, is shaped by perceived usefulness, ease of use, and contextual factors, with mixed attitudes reflecting both enthusiasm and apprehension. Trust also emerges as a critical mediator, with females and lower-proficiency users expressing greater scepticism (Nazaretsky et al., 2025; Saihi et al., 2024). Cultural dimensions further influence perceptions, with calls for ethical guidelines to ensure equitable integration (Yusuf et al., 2024). Student attributes, such as gender, age, and institutional context, also moderate AI-related perceptions, with older students (e.g., in their 30s-40s) and non-native English speakers reporting higher usage for specific tasks (Baek et al., 2024). The adoption of AI is also shaped by students perceived competence when using AI, with a high correlation between AI competence and perceived benefits of AI-tools on the intended use of AI technologies (Delcker et al., 2024). Different frameworks for AI competency have been developed to evaluate students' AI-competence, some of them extending beyond knowledge to also include confidence and self-reflective mindsets (Chiu et al., 2024). Systematic reviews reveal that students often overestimate their digital competences, with gaps in areas like information literacy, content creation, and ethical use (Sánchez-Caballé et al., 2020; Zhao et al., 2021).

However, the interplay between students' willingness to adopt AI companions and their motivations for engaging in traditional academic activities remains a critical yet underexplored area. Self-determination theory (SDT; Deci & Ryan, 2012) provides a robust framework for understanding this dynamic, suggesting that motivation ranges from intrinsic (driven by personal interest) to extrinsic (influenced by external pressures), with the satisfaction of psychological needs, competence, autonomy, and relatedness, playing a pivotal role (Li et al., 2025a). Some research findings indicate that students' motivation to use AI tools often leans toward introjected regulation (e.g., avoiding guilt or shame), with intrinsic motivation less central, implying that AI engagement may stem more from obligation than enjoyment (Li et al., 2025b). Competence satisfaction appears to drive AI adoption more strongly than autonomy or relatedness, suggesting that building students' confidence in using AI tools is crucial for sustained engagement (Ma & Chen, 2025; Xia et al., 2022).

In this sense, the adoption of AI companions may influence motivations for traditional academic activities in complex ways. For instance, by addressing informational needs remotely, AI tools could reduce the perceived necessity of attending lectures or visiting libraries, potentially weakening the communal aspects of university life (Chan & Tsi, 2024). Conversely, by alleviating cognitive burdens, AI companions might enhance students' capacity to engage in face-to-face interactions, such as seminars or peer collaborations, thereby reinforcing the value of the physical campus (Smith, 2023). However, risks such as over-reliance on AI or diminished critical thinking highlight the need for teacher scaffolding to maintain a balance between technological and human elements in education (Ma & Chen, 2025). Addressing these dynamics requires further investigation to ensure that AI companions complement, rather than undermine, the motivational drivers of traditional academic engagement.

## 3. Methodology

To investigate university students' experiences, attitudes, and motivations regarding the use of artificial intelligence (AI) in higher education, this study used a cross-sectional survey design collecting data from a large and diverse student population.

### 3.1. Sample

To ensure the reliability of the dataset, responses that were incomplete or completed in less than five minutes were excluded from the analysis, reducing the sample from 1310 to 926 students from the University of Zurich (UZH). Participants had an average age of 26 (SD = 5.84) and had been studying at UZH for an average of 3 years (SD = 2.61). It should be noted that 36% of respondents had previously studied at another university for an average of 4.4 years. In terms of gender distribution, 58.1% of the sample were women, 38.3% were men, 2.2% identified as non-binary, 1.4% preferred not to disclose their gender. The sample included students from all seven faculties of the university, with the highest proportions from the Faculty of Arts and Social Sciences (38.0%), the Faculty of Science (22.9%), and the Faculty of Business, Economics and Informatics (15.0%). The lowest proportions came from the Faculty of Law (10.5%), the Faculty of Medicine (10.3%), the Faculty of Theology and Religious Studies (1.5%) and the Vetsuisse Faculty, referring to Veterinary Medicine (1.5%). Another 0.3% of respondents selected "Other".

Compared to the total student population of the University of Zurich (UZH) in 2024, which comprised 28.476 students, the current sample of 926 students represented approximately 3.3 % of the total student population. The gender distribution of the sample (58.1% female, 38.3% male, 2.2% non-binary, 1.4% undisclosed) was comparable to the official figures (59.2 % female, 40.8 % male). All seven faculties were represented; however, the sample contained a higher proportion of students from the Faculty of Arts and Social Sciences (38.0% vs. 34.2% of the UZH population) and the Faculty of Science (22.9% vs. 18.1%), and a lower proportion from the Faculties of Medicine (10.3% vs. 14.6%), Law (10.5% vs. 13.8%), and Vetsuisse (1.5% vs. 2.7%).

### 3.2. Instrument

The survey used for this study was developed by a multidisciplinary team of PhD students enrolled in a doctoral excellence programme within a digital initiative of the university. The proximity of the students to the target population (university students) and the diversity of their disciplinary backgrounds ensured a thorough understanding of the research context. The development process began with the identification of five main thematic blocks: demographics, digital literacy, AI skills, the AI companion, and the impact of digitalization on university attendance. The survey was developed and tested in both German and English to accommodate the linguistic diversity of the participants. The survey underwent pre-testing to assess its clarity, comprehensibility, and response time. Feedback from the pre-testing led to further improvements, reducing the number of questions to optimise participant engagement and maintain the validity of the content. The final survey incorporated a variety of question types, including Likert scales, multiple-choice questions and open-ended fields.

The demographic section collected basic information such as participants' year of birth, gender, mother tongue, faculty, programme of study, field of study, and any disabilities. Other questions contextualized participants' academic experience, asking them about their years of study at their current university and at previous institutions. The digital literacy section assessed participants' familiarity with digital tools, their experience with AI technologies, and their use of specific tools for academic, personal, and professional purposes. Their interest in emerging technologies, their proficiency with communication and collaboration platforms, and their knowledge of university guidelines on the use of AI were also explored. The AI skills section assessed participants' perspectives on AI education, including their interest in additional AI courses, the skills they considered necessary, self-assessed competences, and reliance on AI-related resources. The integration of generative AI in their courses and motivations for pursuing or avoiding additional AI education were also investigated. The section on AI-assisted learning examined participants' interactions with AI-based tools and their perceived impact on academic performance. Finally, the section on the impact of digitalization explored how digital technologies influenced participants' motivation to attend university in person and their preferences in a hypothetical virtual reality university environment.

### 3.3. Procedure and data analysis

The data for this study was collected in April 2024. The survey, hosted on Qualtrics, was distributed via the university's email lists to ensure wide dissemination across all faculties. To encourage participation, respondents were offered the opportunity to participate in a draw for 20 vouchers, each worth 50 CHF, redeemable at an online store specializing in digital products. The survey took approximately 15-20 minutes to complete. Before participating, informed consent was obtained from participants, who received clear information about their rights, the use of their data, and the purpose of the study. According to the regulations of the University of Zurich, no formal approval by the ethics committee was required for this study.

Quantitative data were analysed using Jamovi (version 2.3.21.0), a statistical software package based on R. Descriptive statistics, including means, standard deviations, and frequency distributions, were calculated to summarize digital literacy variables. Inferential statistical techniques, including chi-square tests and analysis of variance (ANOVA), were used to examine differences between demographic variables such as gender and faculty affiliation. Correlation analyses were performed to explore relationships between key variables, with a significance threshold of $p < 0.05$. Effect sizes were interpreted using Pearson's r, categorized as small ($r = 0.1$), medium ($r = 0.3$), or large ($r = 0.5$), to quantify the strength of associations. Qualitative data from the open-ended survey responses were analysed using R, leveraging its text analysis capabilities to systematically identify recurring themes and patterns. This dual analytical strategy provided a nuanced understanding of quantitative and qualitative trends, offering a holistic view of the study results.

## 4. Results

### 4.1. AI prior experiences, perceived competence, and interest

This section addresses the first research question: *What are students' prior experiences with AI tools, their perceived digital and AI-related competences, and their interest in emerging technologies*?

Most students reported having considerable previous experience with AI technologies. Specifically, 96.9% of participants indicated that they had already interacted with AI tools, while only 2.4% said they had no experience, and 0.8% were unsure. Among those with experience, AI was most frequently used for study-related purposes (82.4%), closely followed by personal use (76.8%), and a smaller proportion (40.4%) reported using it in professional or work-related contexts. Frequency patterns varied, although a significant proportion of students reported regularly integrating AI into their routines (see Figure 1). When looking for help regarding the use of AI, findings indicate that students mainly turn to the Internet, including online resources and tutorials (89.3%), followed by their fellow students (48.3%), with considerably fewer turning to faculty members or instructors (14.3%), specific courses (5.7%), or central services (5.1.%) at the university, or other resources (6.6%) such as other people or AI itself.

Figure 1. Frequency of use of AI

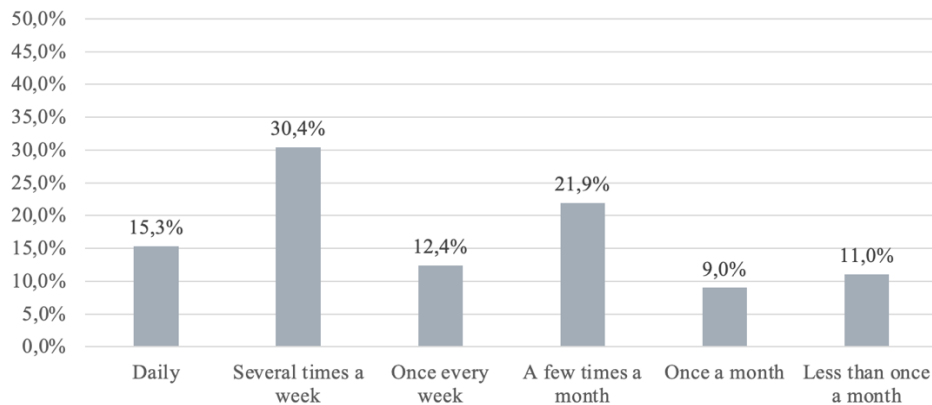

Overall familiarity with digital communication and collaboration platforms was rated relatively high, with an average score of 4.04 (SD = 0.76) on a 5-point scale, suggesting that, in general, students feel confident navigating digital environments. When asked about specific tools, participants most frequently mentioned ChatGPT (835 mentions), followed by DeepL (713), indicating a strong preference for text generation and translation applications. Additional commonly used tools included Grammarly (340), DALL-E (192), and BingChat (161). Other tools, such as GitHub Copilot, MidJourney, and Gemini/Google Bard, were also highlighted. Although used less frequently, tools such as Research Rabbit, Stable Diffusion, and Perplexity illustrated the diversity of AI applications explored by students. In open-ended responses, participants mentioned tools such as Scite or Elicit, among others.

Then, students were asked to rate the usefulness of AI tools for a range of academic and cognitive using a 4-point scale (1 = not helpful at all, 4 = very helpful; see Figure 2). The results suggest that students primarily value AI for text-centric and knowledge-related tasks, while they are more cautious or less interested in its use in technical or visual domains.

Figure 2. Usefulness of AI in the different tasks

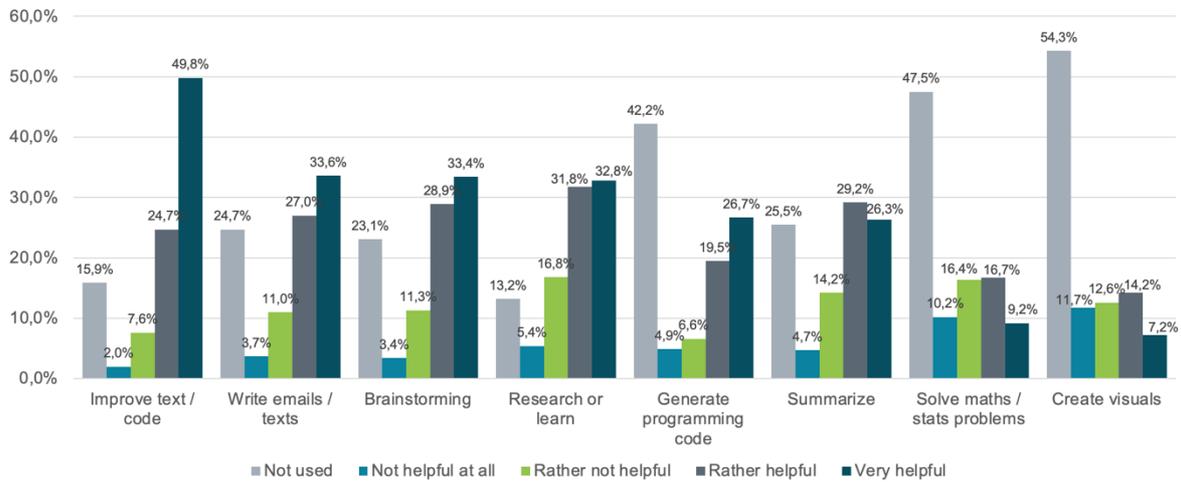

Self-assessed digital competence was measured using a 7-point scale, on which students rated their overall ability to use digital tools with an average of 4.24 (SD = 1.43). To gain a deeper insight into students' perceptions of relevant AI-related skills, participants were asked to choose and rank a list of eight competences in order of importance, many of which were adapted from the DigComp framework (Vuorikari, et al., 2022). Almost all participants considered "critical thinking for answer evaluation" an important skill to have (see Figure 3). And when it comes to the ranked position of each skill, Critical thinking emerged again as the most highly valued skill, with 57.7% of students ranking it as the most important. Prompt writing was also considered a priority, receiving the second-highest proportion of top rankings (19.1% ranked it first and 29.5% second), highlighting students' recognition of the practical skill needed to interact effectively with generative AI systems. In contrast, more technical skills, such as programming knowledge and AI software fundamentals, generally scored lower, with only 8.5% and 14.2% respectively assigning them the highest rank, respectively. Communication skills, knowledge of copyright and licensing, privacy and data protection, and problem solving typically ranked in the middle, suggesting that students consider them relevant but not essential.

Figure 3. Skills considered important when using AI

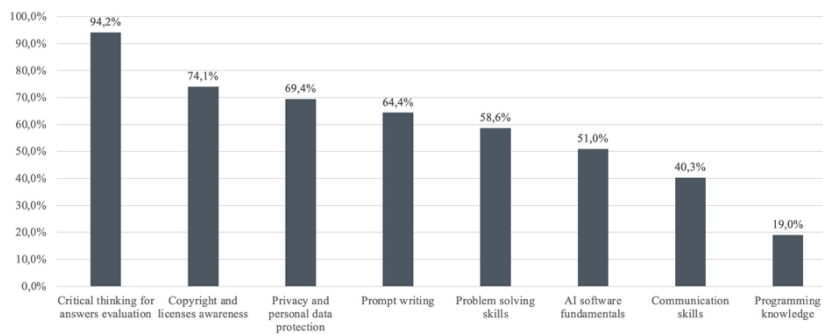

To assess their level of competence in these AI-related skills, students were also asked to assess their own skill level on a 3-point scale (1 = basic, 2 = intermediate, 3 = advanced; see Figure 4). The results reveal that critical thinking was the area in which students felt most confident: 62.3% rated themselves as advanced and only 14.9% as basic. Prompt writing was the second highest rated skill, with 34.4% rating it as advanced and 47.1% as intermediate. In contrast, technical and data-related skills, such as knowledge of AI software and programming, were among the lowest rated. In terms of AI software knowledge, only 5.3% rated themselves as advanced, while almost half (49.5%) said they had basic proficiency. A similar pattern was observed in programming, where 64.5% rated themselves as basic and only 3.3% as advanced. Knowledge of copyright and licensing, as well as privacy and data protection, also received modest ratings, with less than 10% rating themselves as advanced in any of these areas. Communication and problem-solving skills, although not rated as highly as critical thinking, showed stronger perceived competence: 40.9% and 29.0% rated themselves as advanced in these areas, respectively.

Figure 4. Self-assessed level of competence in AI-related skills (1 = basic, 2 = intermediate, 3 = advanced)

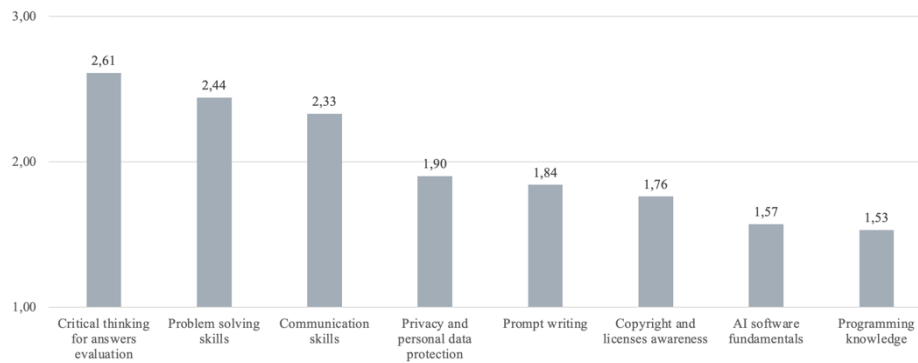

Finally, students' interest in emerging technologies was measured using a 5-point scale, with an average score of 3.48 (SD = 0.97), indicating a moderately high level of interest. When asked about the desirability of offering more AI-related courses, 38.8% of students expressed a desire for more, with common themes including the ethical and social implications of AI, critical engagement strategies, applications for research and education, and data security. In contrast, 9.3% explicitly opposed expanding AI curricula, citing concerns about exaggerated expectations, sufficient exposure, or a preference for developing human-centred skills. A significant proportion remained undecided: 13.1% were unsure about the need for more courses, while 38.9% said they were unaware of any such courses.

Gender differences with statistically significant effects were found in self-assessed digital competence ($F(4, 12.1) = 17.75$, $p < 0.001$) and interest in emerging technologies ($F(4, 12.1) = 8.47$, $p = 0.002$), however the limited sample size of the participants not willing to disclose their gender warrants caution in interpretation. On the one hand, in terms of self-assessed digital competence, descriptive statistics showed that male participants reported the highest average competence (M = 4.74, SD = 1.33), followed by those who preferred not to reveal their identity (M = 4.67, SD = 1.53), and those who identified as non-binary (M = 4.20, SD = 1.64). Female participants had the lowest mean score (M = 3.90, SD = 1.38), suggesting a gender gap in self-assessed digital competence. On the other hand, in terms of interest in emerging technologies, those who preferred not to disclose their gender expressed the highest interest (M = 4.33, SD = 1.16), followed by male participants (M = 3.71, SD = 0.97). Female participants (M = 3.34, SD = 0.93) and non-binary students (M = 3.05, SD = 1.15) had lower levels of interest. No statistically significant differences were found between genders in terms of familiarity with digital platforms ($p = 0.061$) or frequency of use of digital tools ($p = 0.171$).

Statistically significant differences were also observed between faculties in terms of self-assessed digital competence ($F(7, 31.2) = 8.61$, $p < 0.001$), familiarity with digital communication and collaboration platforms ($F(7, 31.1) = 2.91$, $p = 0.018$), interest in emerging technologies ($F(7, 31.4) = 5.35$, $p < 0.001$), and frequency of use of digital tools ($F(7, 31.8) = 6.74$, $p < 0.001$). Specifically, students from the Faculty of Business, Economics and Informatics consistently obtained the highest scores on all four indicators: self-assessed digital competence (M = 4.88, SD = 1.44), familiarity (M = 4.22, SD = 0.70), interest in emerging technologies (M = 3.90, SD = 0.94), and frequency of use (M = 4.46, SD = 1.45). In contrast, students from the Faculty of Theology and Religious Studies and the Vetsuisse Faculty obtained the lowest scores. Theology students rated their digital competence with an average of 3.21 (SD = 1.37), their familiarity with an average of 3.93 (SD = 0.83), their interest in emerging technologies with an average of 3.43 (SD = 0.85), and their frequency of use with an average of 3.29 (SD = 1.59). Similarly, students from the Vetsuisse Faculty rated their competence with an average of 3.57 (SD = 1.34), their familiarity with an average of 3.64 (SD = 0.76), their interest with an average of 3.07 (SD = 0.62), and their frequency of use with an average of 2.62 (SD = 1.32).

Overall, these results reveal a student population that is committed to digital technology, with extensive exposure to AI tools and a strong emphasis on text-based and productivity-enhancing applications. Students demonstrate high self-confidence in their overall digital skills and critical thinking, especially in academic contexts. However, significant gender and faculty disparities highlight uneven digital confidence and unequal interest in emerging technologies.

### 4.2. Attitudes towards an AI Buddy

In this section the second research question is addressed: *How do students perceive a hypothetical "AI Buddy" (a digital companion designed to support students throughout their academic journey) including adoption, benefits and concerns?*

Students' willingness to use the AI Buddy was assessed on a 7-point Likert scale (1 = very unlikely, 7 = very likely; see Figure 5), and the mean score was 4.79 (SD = 1.86) indicating a generally positive inclination toward use. The intended uses of the AI Buddy were also explored (see Figure 6).

Figure 5. Willingness to adopt an AI Buddy

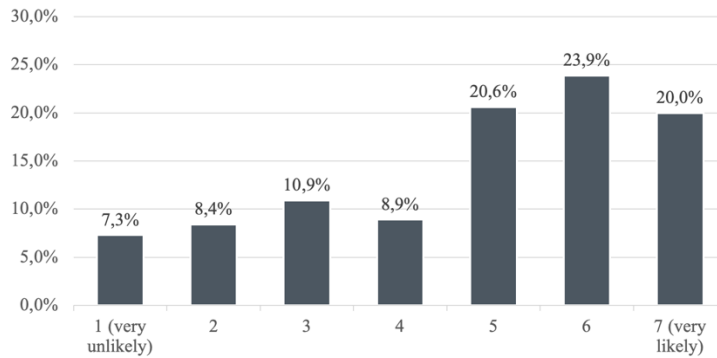

Figure 6. Intended uses of the AI Buddy

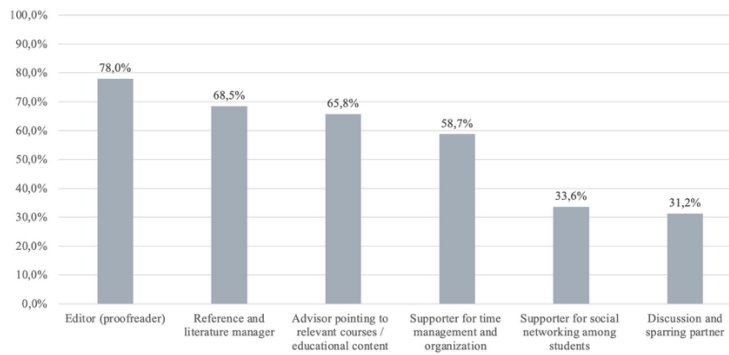

Participants also listed expected benefits and concerns related to the AI Buddy. Benefits were analysed using automated content analysis. The most frequently mentioned benefit-related term was "time" (168 mentions), suggesting that students valued time efficiency. Other prominent terms included "information" (125), "help" (116), "better" (115), "study" (96), "students" (93), "easier" (89), "support" (80), "learning" (79), and "planning" (74). These keywords point to perceived improvements in academic efficiency, ease of learning, and study organization. In contrast, concerns were dominated by issues related to data protection. The term "data" was the most frequent (508 mentions), followed by "protection" (248), "privacy" (124), and "information" (118). Other concern-related terms included "personal" (88), "students" (74), "use" (69), "security" (49), "thinking" (44), and "answers" (42). These terms indicate that data security, personal privacy, and trust in the AI Buddy's handling of information were central to participants' apprehensions.

Also, the results indicate that students are generally selective with regards to the type of information they are willing to share and to whom access should be granted. The majority expressed readiness to share their academic schedule (80.0%), followed by learning preferences (71.0%), academic interests and career goals (69.4%), and study materials (67.6%). More sensitive information, such as academic records (52.5%), personal schedules (34.6%), browsing history (12.2%), and trace data (8.7%), was less likely to be shared. In terms of access, students preferred to retain control over their data, with 91.4% indicating access should remain with themselves. Nonetheless, a considerable proportion was open to granting access to the AI Buddy development team (69.3%) and researchers for academic purposes (57.1%). Access to authorized university officials for administrative purposes (38.7%), other students for research purposes (31.0%), and university officials for assessment purposes (23.2%) was less frequently supported. Only a minority endorsed data access by contracted third-party companies for research (19.7%) or external companies (5.0%). These findings highlight students' willingness to share data that supports their learning and career development, while demonstrating caution with more personal information and external entities.

Finally, participants were asked whether they believed users of an AI Buddy would be advantaged compared to non-users. Responses were relatively evenly distributed: 27% believed users would be advantaged, 28% did not think so, and 45% were unsure. When asked whether they would feel compelled to adopt an AI Buddy if others were gaining an advantage, 58% answered affirmatively, 24% responded negatively, and 18% were unsure. These findings suggest that while many students remain uncertain about the actual benefits of using an AI Buddy, a substantial proportion would be inclined to adopt such a tool to avoid potential disadvantages relative to their peers.

Statistically significant differences in willingness to adopt Buddy AI were also observed by gender ($F(4, 12.7) = 6.73$, $p = 0.004$). Female ($M = 4.83$, $SD = 1.81$) and male ($M = 4.81$, $SD = 1.89$) participants reported very similar levels of willingness, both above the midpoint of the scale. In contrast, non-binary participants ($M = 3.60$, $SD = 2.09$) and those who preferred not to disclose their gender ($M = 3.70$, $SD = 1.83$) reported significantly lower willingness. The highest mean score was observed among the small group of participants who preferred to self-describe their gender ($M = 6.33$, $SD = 0.58$), although the limited sample size ($n = 3$) warrants caution in interpretation.

The differences between faculties in terms of willingness to use AI Buddy were also statistically significant ($F(7, 31.3) = 2.55$, $p = 0.034$). Students in the Faculty of Business, Economics, and Informatics reported the highest mean likelihood of use ($M = 5.29$) followed by the Faculty of Science ($M = 4.85$). In contrast, the lowest mean scores were observed among students in the Faculty of Theology and Religious Studies ($M = 4.21$) and the Vetsuisse Faculty (M = 4.36). Faculties such as Law (M = 4.78), Medicine (M = 4.77), Arts and Social Sciences (M = 4.58) were in the middle range.

### 4.3. AI Buddy and university attendance

In this section the third research question is addressed: *How does willingness to adopt an AI Buddy relate to motivations for engaging in traditional academic activities?*

To examine this, Pearson's correlation coefficient was used to assess the association between the self-reported likelihood of using the AI Buddy and various motivations for attending university. The results reveal nuanced insights into how students' interest in AI-based support tools intersects with their engagement in traditional academic environments (see Table 1).

Table 1. Correlation with Likelihood to use an AI Buddy

| **Motivation for attending university** | **Pearson's r** | **p** |
| --- | --- | --- |
| Visiting lectures | -0.084* | 0.012 |
| Attending tutorial sessions | 0.038 | 0.263 |
| Discussing in seminars | -0.016 | 0.640 |
| Studying in quiet spaces | 0.011 | 0.745 |
| Studying with classmates | 0.034 | 0.310 |
| Socializing with classmates | -0.036 | 0.280 |
| Participating in student events | 0.052 | 0.112 |
| Eating in the cafeteria | -0.031 | 0.352 |
| Using university digital resources (computers, WIFI, etc.) | 0.024 | 0.467 |
| Consulting or borrowing books from the library | -0.105** | 0.001 |

* $p<0.05$, ** $p<0.01$, *** $p<0.001$

A significant but weak negative correlation was found between willingness to use the AI Buddy and motivation to attend lectures (r = –0.084, p = 0.012), suggesting that students more inclined to adopt the AI Buddy were slightly less motivated to participate in lecture-based instruction. This tendency was even more pronounced regarding the use of the university library, with a slightly stronger significant negative correlation between willingness to use the AI Buddy and motivation to consult or borrow books from the library (r = –0.105, p = 0.001). These findings suggest that students who show a higher propensity to engage with AI-based academic support may be somewhat less reliant or perceive less need to be reliant on traditional resources and infrastructures associated with in-person learning. In contrast, no significant associations were found between the likelihood of using the AI Buddy and motivations linked to social or interactive academic experiences. Specifically, motivations such as discussing in seminars (r = –0.016, p = 0.640), socializing with classmates (r =

–0.036, p = 0.280), studying with classmates (r = 0.034, p = 0.310), and participating in student events (r = 0.052, p = 0.112) did not show significant correlations with willingness to use the AI Buddy. Similarly, the motivation to use university digital resources such as computers and Wi-Fi (r = 0.024, p = 0.467), to attend tutorial sessions (r = 0.038, p = 0.263), or to eat in the cafeteria (r = –0.031, p = 0.352) were also not significantly associated with AI Buddy adoption.

Overall, these results suggest that students' willingness to adopt an AI Buddy is modestly related to a reduced inclination toward certain traditional academic activities, particularly lecture attendance and library use. However, it appears largely unrelated to motivations that reflect the social, collaborative, or digitally supported dimensions of university life. This implies that while the AI Buddy may serve as a partial substitute for formal academic structures, it does not replace, nor diminish, the value students place on interactive and communal aspects of the on-campus experience.

## 5. Discussion

This study explored university students' experiences with AI technologies, their appraisal of a hypothetical AI Buddy, and the potential implications of its adoption for campus engagement. The findings offer several important insights for understanding the digital readiness of students, the perceived value and limitations of AI-based academic companions, and their potential impact on students' physical presence in higher education environments.

### 5.1. Students' AI Engagement and Perceived Competence

The results of this study indicate widespread engagement with AI tools among university students, with 96.9% reporting prior use, primarily for academic purposes such as writing, brainstorming, and research, similar to previous studies (Digital Education Council Global AI Student Survey, 2024; Chan & Tsi, 2024). This high adoption rate reflects the rapid integration of generative AI tools into students' routines, consistent with prior findings that highlight their popularity for text-centric tasks (Baek et al., 2024). However, students' self-assessed digital competence and specific AI-related skills reveal significant gaps, particularly in technical areas like programming and AI software knowledge, where only a small part of students rated themselves as advanced, respectively. In contrast, cognitive skills like critical thinking and prompt writing were rated higher, aligning with frameworks that emphasize critical appraisal and practical application as core AI literacy components (Bećirović et al., 2025; Chiu et al., 2024).

Nonetheless, an important finding arises from the discrepancy between the abilities that students regard as important and their self-assessed competency. Students assessed their proficiency in critical thinking and prompt writing as high, and they expressed high levels of overall self-confidence in their digital skills; however, their confidence declined considerably when it came to more technical and legal skills. For example, almost half said they knew "basic" things about AI software, while a much higher percentage said they knew "basic" things about programming. This pattern indicates a notable discrepancy between students' perceptions of what is required for efficient AI application and a more thorough comprehension of the technology. Students see themselves as smart users and evaluators of AI output, confident of their capacity to control the system with well-crafted prompts and to critically examine the outcomes. They do not, however, possess the fundamental technical and legal expertise necessary to interact with AI more deeply, such as comprehending the underlying mechanisms, privacy consequences, or copyright issues. This is consistent with earlier studies showing that students' self-assessed AI competence is linked to their perceived benefits of AI (Delcker et al., 2024) and that perceived AI competency extends beyond real knowledge and is based on confidence (Chiu et al., 2024).

Finally, the statistically significant differences in self-assessed digital literacy and interest in new technologies by gender and faculty affiliation are particularly relevant (Zhao et al., 2021). Male students consistently reported higher levels of competence and interest than their female and non-binary peers, with students from the Faculty of Economics, Business, and Informatics outperforming all other faculties, although this pattern may partly reflect co-variation in gender distribution (e.g., a higher proportion of male students in the Faculty of Economics, Business, and Informatics compared to Vetsuisse). In contrast, students from the Faculty of Theology and Philosophy and the Vetsuisse Faculty scored the lowest. This discrepancy indicates that the benefits of AI competence and implementation are not distributed equally. The self-reinforcing cycle of students in technology-intensive fields having a perceived advantage and greater enthusiasm for new technologies may widen the skills gap with their peers in other disciplines (Baek et al., 2024).

### 5.2. Willingness to Adopt an AI Buddy

Students displayed a generally positive willingness to adopt an AI Buddy, with almost half the sample indicating high likelihood to use it. Preferred functions included editing/proofreading, reference management, and academic advising, aligning with perceived benefits of time efficiency, study support, and learning enhancement (Chan & Hu, 2023; Rahman et al., 2025). Rather than seeing the AI as a companion for socio-emotional needs, which some research suggests can mitigate loneliness (Lai et al., 2025), students perceive it as an efficient, personalized academic assistant. Their primary interest lies in the tool's ability to enhance their productivity and streamline academic tasks. This indicates that students value the AI for its potential to improve their competence and autonomy, consistent with the principles of Self-Determination Theory (Deci & Ryan, 2012), but not for its capacity to foster relatedness or address socio-emotional needs (Li et al., 2025). Moreover, while students generally perceived AI as helpful for certain cognitive tasks, the relatively low scores on perceived usefulness in domains such as mathematics or coding suggest that the AI Buddy's perceived value remains domain-specific and is currently biased towards language-intensive functions. Note that given the rapid development of generative AI's capabilities in these domains (see, e.g., Ziegler et al., 2024 for a study on the use of GitHub copilot by developers), this might have been a function of when the survey was run. Another important aspect arising from the responses is that the AI Buddy should not play a role in assessing or grading students, as the majority were reluctant to grant access to university administrators or lecturers. This highlights a tension with the broader aim of personalized learning: while lecturers might benefit from being informed about individual learning needs for didactic reasons, students clearly resist exposing their individual weaknesses through such tools. This raises important ethical and pedagogical questions regarding the balance between personalization, privacy, and the institutional use of AI in higher education.

However, data protection and privacy concerns took centre stage, reflecting general fears about trust and ethical issues (Yusuf et al., 2024; Nazaretsky et al., 2025). This widespread concern is not a minor obstacle but indicates a potential trust deficit regarding the handling of sensitive personal and academic data by institutions or third-party developers. This finding confirms and emphasises the ethical and privacy concerns expressed in the wider literature on AI integration (Chan & Hu, 2023; Chan & Tsi, 2024). The fact that students' main concerns relate to privacy rather than academic integrity is an important point for higher education policy and AI tool development and emphasises the need for robust and transparent data governance frameworks.

The perception that AI Buddy users might gain an advantage and the pressure to adopt to avoid disadvantage highlight a competitive dynamic that could drive uptake, even among sceptical students (Baek et al., 2024; Li et al., 2025b). However, the lower endorsement for social networking and discussion roles suggests students see AI Buddies as academic tools rather than social substitutes, reinforcing their potential to complement, not replace, human interactions (Chan & Tsi, 2024). Statistically significant differences in the willingness to adopt an AI Buddy were observed across disciplines, with students in the Faculty of Business, Economics and Informatics reporting the highest likelihood of use and those in the Faculty of Theology and Religious Studies reporting the lowest. These differences highlight that openness to AI is not universal and is influenced by both disciplinary context and self-identity, which may require targeted, context-specific institutional strategies for implementation.

### 5.3. Relationship to Physical Campus Engagement

The findings of this study provide insights into how students' interest in an AI Buddy relates to their engagement in traditional academic environments. A significant but weak negative correlation was found between the likelihood of using the AI Buddy and the motivation to attend lectures, a trend that was even more pronounced with the motivation to use the university library. These findings suggest shift in how students plan to engage with academic resources in the situation of having an AI Buddy, considering it a viable alternative for specific, information-heavy functions. An AI Buddy that can summarize articles and manage references might reduce the perceived necessity of visiting a physical library. Similarly, an AI that provides personalized instruction or content might slightly reduce the perceived value of attending a large lecture. This directly addresses the debate about whether the growing reliance on intelligent systems risks eroding the communal and dialogic aspects of university life (Jensen et al., 2025; McGrath et al., 2025). Note, however, that it is likely that both form and content of university classes may change in reaction to the wide-spread availability of AI tools, which may make them more attractive.

An even more interesting finding of this section is the absence of any significant correlation between the willingness to adopt the AI Buddy and motivations for social or collaborative activities. This includes discussing in seminars, studying with classmates, socializing, or participating in student events. This lack of association is a powerful counter-narrative to the idea that AI will erase the university experience. The AI Buddy is not perceived as a substitute for human interaction, peer collaboration, or the communal aspects of university life (Chan & Tsi, 2024). Students appear to maintain a clear separation between the functions of AI (efficiency, information processing) and the functions of the physical university campus (socialization, embodied interaction, and relatedness). This is crucial because it suggests that the core value proposition of the physical university, the satisfaction of the psychological need for relatedness and the creation of a communal, dialogic space, is not threatened by the introduction of an AI companion. This finding directly supports the framework of Self-Determination Theory (Deci & Ryan, 2012), stating that while AI can enhance competence and autonomy, it does not yet supplant the fundamental human need for relatedness (Ma & Chen, 2025).

### 5.4. **Practical Considerations for AI Integration in Higher Education**

Overall, the findings of this study highlight the need for a nuanced approach to integrating AI companions into higher education, balancing technological innovation with ethical, pedagogical and privacy considerations. To fill gaps in technical and legal AI skills, universities should prioritise comprehensive AI literacy programmes tailored specially to those students in less technology-oriented faculties, such as humanities or social sciences, where familiarity with AI tools may be limited. Integrating hands-on training in prompt engineering and critical thinking into curricula can build on students' strengths in analytical reasoning. Workshops, interdisciplinary courses and hands-on simulations of AI tools can build confidence and competence, reducing disparities in AI competence across academic disciplines. For example, case-based learning modules could simulate real-world AI applications, encouraging students to overcome ethical and technical challenges collaboratively.

Furthermore, the students' strong privacy concerns highlight the importance of rigorous data governance in the implementation of an AI Buddy, which should be explicitly excluded from assessing or grading student performance to mitigate risks of data misuse, bias, or breach of trust. University administrators and professors should be restricted from accessing personalized AI Buddy interactions, such as individual learning patterns or academic weaknesses, to protect student privacy and autonomy. This restriction is particularly important in the context of assessment: if AI Buddies were to inform grading or performance evaluations, it could compromise fairness and transparency, reinforce biases, and discourage students from using the tool openly. Instead, AI systems should provide aggregated and anonymized information to inform teaching strategies without compromising individual privacy. For example, anonymized data could reveal common learning challenges across a cohort, allowing teachers to adjust course content without accessing sensitive personal details. This approach aligns with ethical guidelines, such as those of the IEEE Global Initiative on the Ethics of Autonomous Systems, which advocate transparency and minimal data exposure in AI educational tools.

In relation to the role that AI Buddies should occupy, they should be positioned as complementary tools that enhance, not replace, traditional academic activities. Their role in streamlining tasks, such as summarising lecture content, providing research assistance or generating practice questions, can improve efficiency while preserving the social and collaborative aspects of learning. Universities should integrate AI Buddies into curricula in ways that reinforce critical engagement with primary sources, such as lectures and library resources. Faculty training in AI scaffolding techniques is essential to guide students in the critical use of AI tools, ensuring that they do not become overly dependent on automated outputs.

This study also reveals a structural gap between institutional support and student-led learning networks. While students express a clear interest in expanding their AI-related learning, their development currently depends largely on informal exploration rather than structured educational offerings. Students primarily seek help regarding AI use from the Internet and their peers, with a striking lack of reliance on faculty or formal university courses and services. They are not waiting for the university to provide formal training on AI; they have already become self-directed learners in this domain. This dynamic undermines the traditional role of the university as the sole deliverer of knowledge and formal training. It suggests that future university-led AI literacy initiatives must shift from being foundational "how-to" courses to more advanced, collaborative spaces that address the ethical, social, and critical dimensions of AI. Faculty, instead of serving as the primary source of instruction, must be repositioned as facilitators and ethical guides in this rapidly evolving landscape.

Finally, to maximise the educational benefits of AI Buddies, universities could adopt strategies based on the Self-Determination Theory to nurture students' intrinsic motivation. AI Buddies can support autonomy by offering customisable learning pathways, competence by providing adaptive feedback, and relatedness by facilitating peer-like interactions. However, over-reliance on AI could undermine the dynamics of social learning. To counter this, AI Peer interactions should be designed to encourage collaborative activities, such as group problem-solving prompts or peer review facilitated by AI tools.

## 6. Limitations and future directions

This study provides valuable insights into the potential integration of AI Buddies in higher education, but several limitations must be acknowledged to contextualize its findings. The reliance on self-reported data introduces risks of bias, such as social desirability or inaccurate self-assessment, particularly regarding students perceived digital competence. These biases may skew reported proficiency levels, potentially overestimating or underestimating actual AI-related skills. Additionally, the correlational design further restricts causal inferences, and the modest effect sizes observed suggest that findings should be interpreted with caution, as they may not fully reflect the practical significance of AI Buddy adoption. Third, the study's scope was limited to a single institutional context, which may constrain the generalizability of results across diverse academic or cultural settings. Fourth, the student's answers were clearly shaped by the capabilities of today's AI. They do not extend to possible new (applications of) technologies that may change the way we act in academe. And finally, our study design posed many questions in isolation and not in terms of trade-offs. For instance, the quality of certain functionality typically comes at the price of giving up some privacy, a trade-off that many users already engage with in the private sector.

To address these limitations, future research should prioritize longitudinal studies to examine the long-term impacts of AI Buddy integration on academic performance, study behaviours, and campus engagement. Such designs would enable researchers to track how sustained interaction with AI tools influences learning outcomes and student motivation over time, providing a clearer picture of their efficacy and challenges. Qualitative approaches, such as in-depth interviews or focus groups, could offer richer insights into how students navigate the interplay between human and AI-driven support, particularly in terms of trust, autonomy, and perceived value. These studies could explore students lived experiences, shedding light on the emotional and cognitive dimensions of AI Buddy use. Cross-institutional and cross-cultural comparative studies are also needed to enhance the generalizability of findings. By examining AI Buddy adoption across diverse academic environments and cultural contexts, researchers can identify institutional factors, such as infrastructure, faculty training, or student demographics, that shape perceptions and implementation outcomes. Such studies could reveal how varying educational policies or cultural attitudes toward technology influence AI adoption, informing more adaptable and inclusive strategies.

Further research should also investigate the role of sustained AI literacy training in building competency across diverse student populations, particularly those in non-technical disciplines. Theory-based interventions, such as those grounded in Self-Determination Theory, could be tested to assess their effectiveness in fostering intrinsic motivation within AI-supported learning environments. These interventions might focus on designing AI Buddies to support autonomy through customizable learning pathways, competence through adaptive feedback, and relatedness through features that encourage collaborative learning.

By addressing these research gaps, future studies can provide a more comprehensive understanding of AI Buddy adoption in higher education.

## 7. Declarations

The survey instrument and the corresponding dataset generated and analysed during this study are available from the corresponding author upon reasonable request. No funding was received for conducting this study.

The authors gratefully acknowledge the contributions of the doctoral students who participated in the Strategy Lab at the Digital Society Initiative of the University of Zurich, whose interdisciplinary perspectives and commitment were instrumental in the development of the survey instrument: (names). We also extend our sincere thanks to all university student participants who generously shared their time and perspectives by completing the survey.